\documentclass[preprint, showpacs]{revtex4}
\usepackage{amsmath}
\usepackage{amssymb}
\usepackage{graphicx}
\usepackage{bm}
\usepackage{dcolumn}

\begin{document}

\title{Strong electron-lattice coupling as the mechanism behind charge 
density wave transformations in transition-metal-dichalcogenides}

\author{Lev P. Gor'kov}
\email{gorkov@magnet.fsu.edu}
\affiliation{National High Magnetic Field Laboratory, Florida State University, 
Tallahassee, FL 32310}

\begin{abstract}
We consider single band of conduction electrons interacting with 
displacements of the transitional ions. In the classical regime strong 
enough coupling transforms the harmonic elastic energy for an ion to the one 
of the well with two deep minima, so that the system is described in terms 
of Ising spins. Inter-site interactions via the exchange by electrons order 
spins at lower temperatures. Extension to the quantum regime is discussed. 
Below the CDW-transition the energy spectrum of electrons remains metallic 
because the structural vector $\bm{Q}$ and the FS sizes are not 
related. Large values of the CDW gap seen in the tunneling experiments find 
their natural explanation as due to the deep energy minima in the bound 
two-well electron-ion complex. 
\end{abstract}

\maketitle
\section{Introduction}

Origin of charge density waves (CDW) in the transition-metals 
dichalcogenides (TM2HDC) remains enigmatic since their discovery in the 
early 70's \cite{cit1}. At the time, the most popular theoretical 
scenario was the  so-called ``nesting'', the congruency of two or more 
Fermi surfaces (FS) separated by a vector $\bm{Q}$ in 
the momentum space. Interactions with the momentum transfer 
$\bm{Q}$ would lead to the CDW instability 
(with $\bm{Q}$ becoming the structural 
vector) and to the opening of energy gaps on FS. 

The alternative explanation \cite{cit2} ascribes the structural instability to the 
presence of saddle points in the electronic spectrum near the Fermi energy. 
The logarithmic singularities in the electronic density of states (DOS) at 
the saddle points favor instabilities with the momentum 
transfer, $\bm{Q}$ connecting the two saddle points. The 
mechanism also leads to the energy gaps in the electron spectrum. 

Subsequent experiments and band structure calculations for real materials 
gave no support to either of the two concepts. The CDW transition does not 
affect properties of dichalcogenides noticeable. Materials remain metallic 
below the transition temperature, $T_{CDW}$. Superconductivity in 2$H$-TaSe$_2$ and 
2$H$-NbSe$_2$ takes place on the background of the CDW phase. 

The density functional calculations\cite{cit3} (2$H$-NbSe$_2$) have shown 
that the contribution from the nested pieces of FS into the charge 
susceptibility is negligible at the expected value of the vector 
$\bm{Q}$, although, indeed, the ground state of the system at $T = 0$ is the CDW phase.   
In \cite{cit3,cit4} it was hypothesized that CDW transitions in 
2$H$-dichalcogenides are driven by a phonon mode that becomes soft due to 
strong electron-phonon (\textit{e-ph}) interactions, provided, though, that 
the \textit{e-ph} matrix 
element is enhanced at the experimentally known structural vector, 
$\bm{Q}$. 

The microscopic physics \cite{cit3,cit4} explains neither differences between 
properties of 2$H$-TaSe$_2$ and 2$H$-NbSe$_2$, nor the CDW gap 
values $\sim 0.1 eV$ \cite{cit5} which are surprisingly large compared to 
the transition temperatures, $T_{CDW} (Ta) = 122K$ and 
$T_{CDW}(Nb) = 33.5K$. (In the infrared data \cite{cit6} 
(2$H$-TaSe$_2$) the gap of such order of magnitude is noticeable 
even close to the transition temperature). 

The mean field analysis of the CDW--transitions performed in \cite{cit7,cit8} 
has shown that in 2$H$-TaSe$_2$ the coherence length is surprisingly 
short: $\xi_{0} \sim 3-10 \AA$. (The CDW transition in 
2$H$-NbSe$_2$ demonstrates features more typical for a mean field 
structural transition). 

We address these peculiarities below. We argue that one may discern two 
stages at the CDW-formation, at least tentatively. At first, strong \textit{e-ph }interactions 
locally bind electrons and ions together in kind of a polaronic effect which 
was first discussed for V$_3$Si in \cite{cit9}. The potential in which 
moves an ion becomes anharmonic. At a strong enough \textit{e-ph} coupling the potential 
possesses the two deep minima, so that the system can be described in terms 
of Ising spins. The subsequent ordering of the local sites at a lower 
temperature is due to inter-site interactions via the exchange by electrons. 
``Trapping'' of electrons by the elastic field is responsible for the main 
energy gains and determines, whereby, the CDW gap by the order of 
magnitude. The available experiments are discussed below in the light of these 
results. 

\section{Choice of the model.}

Before proceeding further, we briefly summarize the information pertaining 
to the CDW-transitions in TM2HGC. The mean-field parameters governing the 
transition were discussed in \cite{cit7}. The lattice superstructure below 
$T_{CDW}$ \cite{cit1} is formed by the triple-$\bm{Q}$ 
modulations of the ionic positions along the three symmetry axes of the 
hexagonal \textit{2H}-phase. The $\bm{Q}$-vector may be incommensurate, but 
even then $\bm{Q}$ remains very close to $\vec{a}^{\ast}/3$ 
(here $a^{\ast}=4\pi /\sqrt 3 a)$. 

It was perceived already in the early 70's\cite{cit10} that the structural changes 
in 2$H$-NbSe$_2$ are related to displacements of the niobium ions. 
Correspondingly, in what follows, we choose a simplified model and consider 
only shifts of the transition-metals ions (Nb or Ta) along each of these 
three symmetry lines. The dramatic softening of the longitudinal acoustic 
mode with $\bm{Q}$ about $\vec{a}^{\ast}/3$ was recently 
observed in \cite{cit11}. Note, in passing, that the optical modes and the 
longitudinal acoustic branch of the same $\Sigma_1$- symmetry are 
actually linearly coupled near point $\vec{a}^{\ast}/3$ of the Brillouin 
Zone (BZ). We return to this later to show that softening of the optical 
mode is accompanied by softening of the acoustic phonon, and vice versa. 

Transition-metals-dichalcogenides are layered compounds with the weak Van 
der Waals coupling between the layers. We restrict ourselves by the 
quasi-two-dimensional properties (Q2D) of a single layer. 

We choose a single isotropic band of electrons interacting strongly with 
displacements of the transition-metal ions. Implicitly, the band is assumed 
to be predominantly of the 4f- (Nb) or 5f- (Ta) - character.

In the model \cite{cit3,cit4} the ``bare'' frequency of the \textit{propagating phonon mode}, 
$\omega_0^2 (k)$ is renormalized by the polarization operator \cite{cit12} 
(assuming the \textit{e-ph} matrix element $g(p,p+k)$is peaked at 
$\vec{k} \approx  \vec{Q}$). Unlike \cite{cit3, cit4}, our attention is concentrated on the \textit{local 
environment} of a single ion. We assume that the optical phonon frequency has 
a negligible dispersion at $\vec{k}=\vec{Q}$. In this respect, as we 
discuss later, our model shows some features common to the Holstein model\cite{cit13}. 

For a heavy ion one may expect that quantum effects are not crucial for 
properties of the lattice, at least, at not too low a temperature. The 
temperature of the (incommensurate) transition in 2$H$-TaSe$_2$, 
$T_{CDW} = 122 K$ \cite{cit1} is rather high and, in the first 
approximation, one can neglect the kinetic energy of the Ta ions. 
 
With the notations $u_j$ for a displacement of the single ion at the 
point $R_i $ the ``bare'' elastic matrix is: $U(R_i -R_j) u_i u_j$.  
At $i=j$ it is merely the potential energy of an oscillator: 
\begin{equation}
\label{eq1}
U(R_i -R_j )u_i u_j \Rightarrow M \omega_0^2 u_i^2/2 = k u_i^2/2.
\end{equation}
Interactions with electrons change the elastic matrix. We calculate the 
contribution to the elastic energy that is due to interactions of electrons 
with arbitrary static displacements. 
With the Hamiltonian for \textit{e-ph} interaction in the form 
\begin{equation}
\label{eq2}
\hat{H}_{e-ph} = \sum_i g u_i  \hat{\psi}^{+} (R_i) \hat{\psi}(R_i)
\end{equation}
the energy of the coupled electron-lattice system $E(u_1 ... u_N;g)$ can 
be calculated from the equation:  
\begin{equation}
\label{eq3}
\partial E(u_{1} ...u_{N} )/\partial g=\sum_i u_i  n(R_i)
\end{equation}
$n(R_i)= \langle \hat{\psi}^{+}(R_i) \hat{\psi}(R_i) \rangle$ is the number of 
electrons per unit cell at the point $R_i$. In terms of the electronic 
Green function (at $T=0$, \cite{cit12}),  
$G_{\alpha \beta }(R,t;R',t') = -i \langle T(\psi _{\alpha} (R_i,t) \hat{\psi }_{\beta }^{+}(R',t') \rangle$, 
one has:  $n(R_i) = - 2 i G(R,t;R,t+\delta)$. (The Green function is 
diagonal in the spin indices: $G_{\alpha \beta }(R,t;R',t') = -i \delta_{\alpha \beta } 
G(R,t;R',t')$ ).

\section{Coupled electron- lattice system in the classical regime.}

At static displacements, one needs to know only the frequency component, 
$G(R,R';\omega)$: 
\begin{equation}
\label{eq4}
G(R,t;R',t') = \int \frac{d \omega}{2 \pi}\, G(R,R';\omega)\exp(-i \omega (t-t'))
\end{equation}
Re-arrange the power expansion of $G(R,R';\omega)$ in $u_i$ as: 
\begin{widetext} 
\begin{eqnarray}
\label{eq5}
G(R,R';\omega )&=& G_0 (R-R';\omega )+\sum_i G_0(R-R_i;\omega) g \bar{u}_i 
G_0 (R-R_i;\omega) \\ 
&+& \sum_{i \ne k} G_0(R-R_i;\omega) g \bar{u}_i G_0(R_i-R_k;\omega) g \bar{u}_k G_0(R_k - R';\omega)+... \nonumber
\end{eqnarray} 
\end{widetext}
The \textit{e-ph} contributions in all powers in $g$ are now summed first for the single 
site: $u_i \Rightarrow \bar{u}_i$. Each of $\bar{u}_i,\bar{u}_k ... $ in Eq.(\ref{eq5}) 
stands for the ``dressed'' local deformation at the corresponding site, $R_i$, $R_k$,.., and is determined by the relation: 
\begin{equation}
\label{eq6}
\bar{u}_i = u_i + u_i G_0(R_i = R_i;\omega) g\bar{u}_i 
\end{equation}
The free Green function,$G_0 (R_i=R_i;\omega)$\cite{cit12} equals:
\begin{widetext}
\begin{equation}
\label{eq7}
G_0 (R_i =R_i;\omega)=\int \frac{d^2 \bar{p}}{(2\pi)^2}\, 
(\omega - E(p)+ E_F +i sign(\omega) \delta)^{-1}
\end{equation}
\end{widetext}
With the help of identity 
\begin{widetext}
$$(\omega - \xi + i sign(\omega) \delta )^{-1} = P\{\frac{1}{\omega 
-\xi}\,\} -i \pi sign(\omega) \delta(\omega -\xi )$$
\end{widetext}
and assuming the electron-hole symmetry for $|\xi| = |E(p)-E_F| $, one obtains: 
\begin{equation}
\label{eq8}
G_0 (R_i =R_i;\omega) = -i \pi \nu (E_F) sign(\omega) 
\end{equation}
After trivial calculations, one finds the electronic contribution into the 
local elastic matrix at the site $R_i$, $E(u_i) = -2 i \int_0^1 
dg u_i G(R_i =R_i; t=t'+\delta ;g)$:
\begin{equation}
\label{eq9}
E(u_i) = -(W/2\pi)\ln\{1+(\pi \nu (E_F) g u_i)^2\}
\end{equation}
($W$ is the bandwidth). $E(u_i)$ is determined as the energy per one ion,  
$\nu (E_F)$ is the number of states at the Fermi level per unit cell:  
$\nu (E_F)=\left(\frac{m}{2\pi }\,\right) S_0$, where $S_0$ is the area of the 2D unit cell. 
$W \nu (E_F)=s$ is an insignificant model parameter, $s \sim 1$. We take $s=2$.

The interaction between two sites, $E(u_i ,u_k)$, is given by the 
second term in Eq.(\ref{eq5}):
\begin{equation}
\label{eq10}
E(u_i ,u_k)=u_i u_k (-i)g^2 \int \frac{d \omega}{2\pi }\, 
G_0^2 (R_{i,k} ;\omega )
\end{equation} 
($R_{i,k} \equiv |R_i -R_k|)$.

The analytic form of $G_0 (R_{i,k};\omega)$ being cumbersome in 2D, for 
the estimate we use its asymptotic at $p_F R_{i,k} > 1$:
\begin{widetext}
\begin{equation}
\label{eq11}
G_0(R;\omega ) \Rightarrow i \nu(E_F) \sqrt{\frac{2\pi }{p_F R}\,}  
\exp\{i sign(\omega) [(p_F +(\omega /v_F ))R - \pi /4)]\}
\end{equation}
\end{widetext}
to obtain:
\begin{equation}
\label{eq12}
E(u_i,u_k)=u_i u_k g^2 \nu^2(E_F)(p_F v_F) \frac{\sin(2 p_F R)}{(p_F R)^2}\,
\end{equation}
The total elastic energy of a single ion is the sum of Eqs. (\ref{eq1}) and (\ref{eq9}). 
Introducing: 
\begin{equation}
\label{eq13}
g^2=\left(\frac{M\omega_0^2}{2 \pi \nu (E_F)}\, \right) \Lambda^2
\end{equation}
\begin{equation}
\label{eq14}
\frac{\pi }{2}\, (M \omega_0^2 \nu (E_F)) = \frac{1}{u_0^2}\,
\end{equation}
and the dimensionless notations for the ion's displacement, $\tilde{u}_i=(u_i/u_0)$, one  
writes down the local elastic energy in the following simple form:
\begin{equation}
\label{eq15}
U_{tot} (u_i) = (1/\pi \nu (E_F))\{\tilde{u}_i^2 -\ln [1+\Lambda^2\tilde{u}_i^2 ]\}
\end{equation}
There, $\Lambda^2$ is the square of the dimensionless \textit{e-ph} coupling constant 
and $u_0$ determines the spatial scale of the local elastic potential. The 
energy scale $T^{\ast }=1/\pi \nu (E_F) = W/2 \pi$ is expressed through the 
electronic DOS. For 2$H-$NbSe$_2$ the band calculations\cite{cit14} gave  
$\nu (E_{F}) = 2.8 \ states/eV$. Hence, $T^{\ast }$ is of the order of tenths of 1 eV.
 
At $\Lambda^2 > 1$ the potential $U_{tot} (u_i)$ has two deep 
minima at $\tilde{u}_{+,-} = \pm \sqrt {1-\Lambda^{-2}}$: 
\begin{equation}
\label{eq16}
U_{tot} (u_{+,-} )=(1/\pi \nu (E_F))\{1-\Lambda^{-2}-\ln \Lambda^2\}
\end{equation}
At $\Lambda^2 < 1$ and temperatures below $T^{\ast}$ $\tilde{u}_i \ll 1$ and  
$U_{tot}(u_i)$ can  be written as:  
\begin{equation}
\label{eq17}
U_{tot}(u_i)=(1/\pi \nu (E_F)) \{\tilde{u}_i^2 (1-\Lambda^2) + \frac{\Lambda^4}{2}\,  
\tilde{u}_i^4 \}
\end{equation}
The quartic term in Eq. (\ref{eq17}) is small, but the anharmonic contribution into 
the elastic energy is the necessary ingredient in the molecular field 
approach to a CDW transition (see e.g. \cite{cit15}).

Re-writing$(p_F v_F)$ as: $(p_F v_F) = (1/\pi \nu (E_F))(p_F^2 S_0/2)$,  
where $S_0 = (\sqrt{3}/4) a^2$ is the area of the triangular 
unit cell, Eq. (\ref{eq12}) in the same notations of Eqs. (\ref{eq13},\ref{eq14}) takes the form:
\begin{widetext}
\begin{equation}
\label{eq18}
E(u_i ,u_k )=\left(\frac{1}{\pi \nu (E_F )}\, \right) \left\{ \left(\frac{\sqrt{3} }{4\pi^2}\,\right) \Lambda^{2} \tilde{u}_i \tilde{u}_k \left(\frac{a}{R}\,\right)^2 \sin (2p_F R) \right\}
\end{equation}
\end{widetext}
The numerical factor in this expression shows that the inter-site 
interactions are weak compared to the on-site $U_{tot} (u_i)$, so that 
the structural transformation, if any, will occur at $T_{CDW}$ well 
below $T^{\ast}$. At $\Lambda^2 > 1$ the model reduces to the 
model of interacting Ising spins:  
$$\tilde{u}_{+,-} (i)\Rightarrow \sqrt{1-\Lambda^{-2}} \sigma_{i}, \ \ (\sigma_{i} = \pm 1).$$
 
When $\Lambda^{2} < 1$, the expressions for $U_{tot} (u_{i} )$ and Eq. (\ref{eq18})  
for $E(u_{i} ,u_{k})$ together complete formulation of the problem: in 
the classical regime the system is fully described by the partition 
function $Z(T,g)$: 
\begin{equation}
\label{eq19}
Z(T,g)= \int (\Pi du_i ) \exp[-\frac{1}{T}\, \sum_{i,k} U_{tot}(u_i ,u_k)]
\end{equation}
( $U_{total} (u_i ,u_j )= U_{tot} (u_i) + E(u_i ,u_k))$. 

The phase transitions are commonly treated in the molecular field 
approximation (see, e.g., \cite{cit15}). The method is not exact for short-ranged 
interactions, such as $E(u_i ,u_k)$ in Eq. (\ref{eq18}) and we do not stay on 
details of the CDW transitions in TM2HDC. 

For the local properties one has for the partition function, $Z(T,g)$:
\begin{widetext}
\begin{equation}
\label{eq20}
Z_{i} (T,g) = \int {d\tilde{u}_i} \exp [-\frac{1}{\pi \nu (E_F)T}\, 
\{\tilde{u}_i^2 - \ln (1+\Lambda^2 \tilde{u}_i^2)\}]
\end{equation} 
\end{widetext}
Scattering of electrons on the lattice displacements above 
$T_{CDW}$ is characterized by the average $\langle g \bar{u}_i \rangle$ that 
enters the denominator of the Green function:
\begin{equation}
\label{eq21}
G^{-1}(R;\omega ) = \omega - E(p)+ E_F + \langle g \bar{u}_i \rangle
\end{equation}
From Eq. (\ref{eq6}) it follows: 
$g \bar{u}_i = gu_i \{ 1-i sign(\omega) \pi \nu (E_F) g u_i \}^{-1}$.  
In the normal phase terms odd in $u_i$ can be  omitted. In dimensionless 
variables:
\begin{equation}
\label{eq22}
g \bar{u}_i = i sign(\omega) \left(\frac{1}{\pi \nu (E_F)}\, \right)  
\frac{\Lambda^2\tilde{u}_i^2}{1+\Lambda^2 \tilde{u}_i^2}\,
\end{equation} 
At $\Lambda^2 < 1$ the imaginary part in $G^{-1}(R;\omega )$ is 
$\langle g\bar{{u}}_{i} \rangle \cong i sign(\omega) \Lambda^2(\pi /2)\nu (E_F )T$: 
in  the classical regime above $T_{CDW}$ the resistivity of the system would be linear 
in T. In the opposite limit of $\Lambda^2 > 1$,  
$\tilde{u}^2 = \tilde{u}_{+,-}^2 = 1 - \Lambda^{-2}$, and the imaginary part is a constant of 
order of  $ \sim 1/\pi \nu (E_{F})$ . The entropy for the system of non-interacting Ising 
spins is finite. 
 
In the ordered state with all ions occupying same minima, 
non-zero $g\bar{u}_i \ne 0$ stands together with the chemical potential 
by:  
\begin{equation} 
\label{eq23} 
g\bar{u}_i = \pm (1/\pi \nu (E_F)) \sqrt{\Lambda^2-1} 
\end{equation}

\section{Discussion of the results}

\subsection{Classical regime}

It is interesting to discuss relevance of the above to experimental results, 
in particular, for 2$H$-TaSe$_2$.

First, recall that the non-linear potential (\ref{eq9}) was derived for a single 
ion. For Eq.(\ref{eq9}) to be meaningful, $u_0$ must be small compared to the 
lattice parameter, $a \sim 3.45 \AA$. With $\omega_0 \sim 10 meV$ 
taken for niobium, Eq.(\ref{eq14}) gives $u_0 \le 0.5 \AA$. This justifies the 
assumption. Experimentally, shifts of cations 
in the superlattice lie in the interval between $0.1 \AA$ \cite{cit1} and 
$0.5 \AA$ \cite{cit9}, i.e. have same order of magnitude as $u_0$, as it should be 
for the  model of Ising spins.

The CDW-gaps seen in the tunneling non-linear current characteristics, 
$dI/dV$ \cite{cit6} (at $T=4.2K$) were large: $\sim 90 meV$ 
and $\sim 60 meV$ for 2$H$-TaSe$_2$ and 2$H$-NbSe$_2$, respectively. 
Such large gaps admit the interpretation in terms of energy of the bound 
electron-ion complex. As mentioned above, the energy scale, $T^{\ast }=1/\pi \nu (E_F)$ 
lies between 0.1-0.3 eV, depending on the number of states per 
unit cell in the specific material (for NbSe$_2$ $\nu (E_F ) \sim 2.8 \ States/eV$ 
per unit cell \cite{cit14}).The value of $|U_{tot}(u_{+,-} )|$ may be adjusted 
even better to the observable value at  $\Lambda^{2} > 1$.

A broad structure in the infrared conductivity with a threshold at 
$\sim 0.25 eV$ was discernable up to 80K \cite{cit5} (2$H$-TaSe$_2$). In recent 
ARPES experiments \cite{cit16} features related to the CDW gap in 2$H$-NbSe$_2$ were 
observable even in the normal phase at 115K. 

Above $T_{CDW}$ the imaginary part in Eqs. (\ref{eq21}, \ref{eq22}) is large$\sim 1/\pi \nu (E_F)$, 
i.e., formation of the local electron-ion complex 
involves electrons from the considerable part of the band. At $T <  T_{CDW}$ the CDW order parameter 
is proportional to ionic displacements in the 
superlattice $\propto g\bar{u}(Q)$. No nesting is involved and, hence, no 
energy gap appears at the Fermi surface. 

(In the CDW-ground state constructed in \cite{cit17} for the tight-binding model 
with the prevailing hopping matrix elements between the next- nearest 
neighbors, the FS at the $\Gamma$-point of BZ would remain metallic either. 
Note, however, that in \cite{cit17} the CDW transition itself takes place due to 
nesting on other parts of the electronic spectrum). 

The frequency of the collective excitations behaves differently at $T=$ 
$T_{CDW}$ for the local potentials with one or two minima \cite{cit16} (see 
also in \cite{cit15}). In the former case the frequency vanishes at the temperature 
of the transition. In the latter, the frequency remains finite as an ion is 
now ``trapped'' by one of the two minimum. Neutron experiments \cite{cit1}(b) for 
2$H$-TaSe$_2$ gave $\omega^2(Q) \simeq 20(meV)^2$ at $T_{CDW}$. Softening of the 
acoustic phonons in 2$H$-NbSe$_2$ at $T_{CDW}=33.5K$ was observed, 
instead, in \cite{cit11}. 
 
The difference in the phonon modes' behavior in the two materials needs a 
clarification. Recall that the longitudinal acoustic branch is coupled 
linearly with the optical mode, $u$ of the same $\Sigma_1$- symmetry at 
$\bm{Q} =\vec{a}^{\ast}/3$. The Free Energy then has a 
contribution of the form:$F(u,s)=\omega_s^2 (s^2/2)+\omega_u^2 (u^2/2)+ t u s$,  
where $s$ stands for the acoustic branch. Minimizing 
$F(u,s)$ and excluding $u$ gives  
$F(u,s)=[\omega_s^2 -(t^2/\omega_u^2)](s^2/2)$. Let the optical mode,  
$u$ be the mode that drives the transition. At $\omega_u^2 (T)\to 0$  
(one minimum), the acoustic mode is 
the first one that manifests onset of the transition. If the potential has a 
few minima, the effective frequency of the acoustic mode  
$\omega_{s,eff}^2 =[\omega_s^{2} -(t^2/\omega_u^2)]$ at 
$T_{CDW}$  may or may not be zero depending on the temperature  
behavior of $\omega_u^2(T)$. Data \cite{cit1}(b) for 
2$H$- TaSe$_2$ agree better with the second possibility.

The two 2$H-$material have different masses of Ta- and Nb- ions. It is known 
\cite{cit16} that at lower temperatures, when quantum effects become important, 
$\omega_u^2 (T)\to 0$ even for a potential with a few minima. This would 
make possible the vanishing of the acoustic phonon frequency at 
$T_{CDW} = 33.5 K$ in the 2$H$-NbSe$_2$ \cite{cit11} and the finite $\omega^2(Q)$ at
$T_{CDW} = 122 K$ in 2$H$-TaSe$_2$ \cite{cit1}(b). 
 
\subsection{Quantum regime}

At low $T$ quantum effects become important in few aspects. Consider first the 
Schr\"{o}dinger equation for an ion moving in the rigid potential  
$U_{tot}(u_i)=(1/\pi \nu (E_F)) \bar{U}(\tilde{u}_i)$, with 
$\bar{U}(\tilde{u}_i)$ in the dimensionless notations. With the kinetic 
energy in the same notations, one has:
\begin{equation}
\label{eq24}
-\frac{1}{2\bar{M}}\, \frac{d^2}{d \tilde{u}^2}\, \Psi(\tilde{u})+ 
(\bar{U}(\tilde{u})-\bar{E})\Psi(\tilde{u})=0
\end{equation} 
The dimensionless ``mass'', $\bar{M}$ in (\ref{eq24}) is defined 
by $1/\bar{M}=(\pi \nu (E_F )\omega_0)^2/2$. (The adiabatic 
parameter $1/\sqrt{\bar{M}} =(\pi \nu (E_F )\omega_0 )/\sqrt{2} $ is 
about one tenth in 2$H-$NbSe$_2$). 

For the single-minimum well ($\Lambda^2 < 1$) quantum effects are 
important when $T_{CDW} < \omega_0$. The lattice 
oscillations are now quantized and the electronic Green function is 
``dressed'' by phonons. In the adiabatic approximations, solution for the 
problem of interacting electrons and phonons in normal metals was given many 
years ago in \cite{cit18}. 

Eq. (\ref{eq24}) adds a possibility of quantum tunneling between minima of the 
two-well potential. However, Eq.(\ref{eq24}) can only describe tunneling in a 
``rigid'' potential, a potential built only by the interatomic forces in the 
lattice. Meanwhile, the two-well potential $U_{tot} (u_i )$ in 
Eqs.(\ref{eq9},\ref{eq15}) 
comes about selfconsistently due to interactions between electrons and the local lattice 
distortions. In a tunneling event, at which the ion goes, say, from $u_{+}$
to $u_{-}$, the electronic configuration reverses as well. Such a feature 
cannot be described in terms of Eq.(\ref{eq24}). Besides, in the quantum regime 
$U_{tot} (u_{i})$ itself varies with temperature\cite{cit9}.

The collective tunneling was discussed in \cite{cit9} in connection with the 
martensitic transitions in Nb$_3$Sn and V$_3$Si. The 
authors were concerned with its mapping on the Kondo problem. (Another 
analogy would be the two-level quantum system). No closed solution was 
obtained in \cite{cit9} for an effective tunneling matrix element. 

Below the transition in the CDW ground state ions will occupy their proper 
minima, in accordance with the superlattice pattern \cite{cit1}(b) and tunneling 
between minima must stop. Nevertheless, experimentally, the pattern reveals 
an interesting peculiarity. Indeed, in the CDW phase one in three atoms 
along the symmetry lines finds itself in the position with the trigonal 
symmetry (2$H$- TaSe$_2$ \cite{cit1}(b); see Fig.4 in \cite{cit8}). While below transition the 
intersite interactions do indeed arrest quantum tunneling between minima for 
the other two of the three atoms, the degeneracy is not lifted for the atom in 
this symmetric position and the collective quantum tunneling processes 
between minima continue at this site down to $T=0$. 
 
According to ARPES data \cite{cit19}, the CDW transition in 2$H$- NbSe$_2$ slightly 
affects only the two-barrel Fermi surface at the K-point in BZ, while FS at 
the $\Gamma $ - point remains intact. (Three $Q$-vectors couple together the 
three points on the inner FS at K, which is seen experimentally as small but 
observable gaps at these three points. The small local gaps ($ \sim 2.4 meV$) survive even  
in the normal phase at 115K. This result is not quite clear.)

The physics of strong local \textit{e-ph} interaction bring us back to the Holstein model 
\cite{cit13} of electrons interacting with dispersionless phonons. No exact solution 
is known for the Holstein model either, but its low temperature physics was 
investigated numerically in DMFT (Dynamical Mean Field Theory, \cite{cit20}) 
approximation. (For a brief summary of results for the Holstein model at 
T$=$0 see \cite{cit21}).

The DMFT approach is strictly local. Intersite interactions (see Eq.(\ref{eq18})) 
and the CDW transition itself cannot be treated by DMFT.

Here we indicate parallels between our physics above and the results \cite{cit21} at 
T$=$0 for the Holstein model. Among them:

1) The double-well potential develops when the e-ph coupling constant becomes 
larger some critical value; 2) Large imaginary part in the Green function: 
the Spectral Function extends over the energy interval that significantly 
exceeds the phonon frequency (compare with our Eq. (\ref{eq22})); 3) The ground 
state remains metallic (at least for not-too-strong \textit{e-ph} coupling).

In \cite{cit21} the 
mass of electronic excitations increases as the residue at the pole of the electronic Green 
function decreases. Judging by these results, the physics studied above 
allows its extension into the low temperature regime.

\section{Summary}

In summary, strong \textit{e-ph} interactions may change the local environment in the 
lattice qualitatively by binding ions and electrons together, thus breaking 
the adiabatic approximation \cite{cit18}. In such case the CDW transition is the 
phase transition in the system of interacting Ising spins. The transition is the 
result of intersite interactions that come about due to the exchange by 
electrons between local sites. The sizes of the structural vector 
$\bm{Q}$ and FS being not related to each other, the energy 
spectrum of electrons remains metallic below the CDW transition. Large 
values of the CDW gaps observed experimentally receive the natural 
explanation in terms of deep energy minima of the two-well local potential. 
Conclusions from the analytical results derived in the classical regime can 
be extended to lower temperatures as it follows from the comparison with 
numerical results for the Holstein model.

\section{Acknowledgments}

The author thanks T. Egami and V. Dobrosavljevic for helpful discussions. 

The work was supported by the NHMFL through NSF Grant No. DMR-0654118 and 
the State of Florida.

\end{document}